\documentclass[sigconf]{acmart}

\copyrightyear{2026}
\acmYear{2026}
\setcopyright{cc}
\setcctype{by}
\acmConference[ICER 2026 Vol. 1]{Proceedings of the ACM Conference on International Computing Education Research Vol.1}{August 11--14, 2026}{Uppsala, Sweden}
\acmBooktitle{Proceedings of the ACM Conference on International Computing Education Research Vol.1 (ICER 2026 Vol. 1), August 11--14, 2026, Uppsala, Sweden}
\acmDOI{10.1145/3765964.3811653}
\acmISBN{979-8-4007-2203-5/2026/08}

\usepackage{graphicx}
\usepackage{threeparttable}
\usepackage{framed}
\usepackage{enumitem}
\usepackage{csquotes, makecell}

\usepackage[skip=3pt]{caption}

\begin{document}

\title{"Why Put in This Much Effort?": How AI Availability Shapes Students' Motivation in Introductory Programming}

\author{Keith Tran}
\orcid{0009-0006-2591-2227}
\email{ktran24@ncsu.edu}
\affiliation{%
  \institution{North Carolina State University}
  \city{Raleigh}
  \state{NC}
  \country{USA}
}

\author{Colton Harper}
\orcid{0000-0002-4745-7409}
\email{colton@umd.edu}
\affiliation{%
  \institution{University of Maryland}
  \city{College Park}
  \state{MD}
  \country{USA}
}

\author{Thomas Price}
\orcid{0000-0001-9375-2292}
\email{twprice@ncsu.edu}
\affiliation{%
  \institution{North Carolina State University}
  \city{Raleigh}
  \state{NC}
  \country{USA}
}

\renewcommand{\shortauthors}{Tran et al.}

\begin{abstract}

\textbf{Background.}
When AI tools can easily complete programming assignments, students face a motivational question: why invest effort in completing them independently? 
While prior work has examined instructor policies and usage patterns, we focus on how students themselves experience and respond to AI availability, a perspective important for designing courses that sustain engagement with programming practice.

\noindent\textbf{Objectives.}
We investigate two research questions: (1) How do engineering students describe how AI availability shapes their motivation to put effort into programming assignments? (2) How do students navigate the tension between their expressed value for learning through effort and the constant availability of AI as an alternative to effort?

\noindent\textbf{Method.}
We conducted semi-structured interviews with 13 engineering majors in an introductory MATLAB course where students could use a course-specific AI chatbot. Using Situated Expectancy-Value Theory (SEVT) as an analytical framework, we examined how students described their expectancy, values, and costs in the context of AI availability.

\noindent\textbf{Findings.}
When AI could complete assignments quickly, students questioned whether their time on programming was well spent (cost), questioned the long-term usefulness of programming skill (utility value), reported less satisfaction when AI bypassed productive struggle (intrinsic value), and described confidence that depended on AI being available (expectancy). Nearly all students expressed a preference for learning through effort and a simultaneous temptation to take shortcuts with AI (sanctioned or otherwise). Some students managed this tension through reframing and explicit boundaries on AI use. Others used AI in ways they saw as conflicting with their own values.

\noindent\textbf{Implications.} Our findings complicate the assumption that students need external constraints to protect their learning. Most students in our sample already valued independent effort but struggled to act on those values when AI offered a faster alternative. Students who managed the tension found motivation in the learning process itself, suggesting that course design may need to shift from valuing what students produce to supporting how they learn.

\end{abstract}

\ccsdesc[500]{Social and professional topics~CS1}
\ccsdesc[300]{Human-centered computing~Empirical studies in HCI}

\keywords{situated expectancy-value theory, end-user programmers, AI code generation, LLMs, student motivation}

\maketitle

\section{Introduction}
\label{sec:intro}

Learning to program requires sustained effort. Students design solutions, write and test code, trace through logic to find errors, and iterate until something finally clicks. Research on the ``doer effect'' has established that students learn through this kind of active practice~\cite{koedinger2016doer}, and these hands-on experiences are central to how students build genuine skill. Throughout this paper, we refer to generative AI tools that can produce working code from natural language prompts, such as ChatGPT, as \textbf{AI}. These tools can now solve most programming problems students encounter at the introductory level~\cite{savelka2023thrilled, finnie-ansley2023my, bernstein2025beyond}, and when a working solution is available on demand, the value of sustained effort may feel less certain to students. Emerging evidence suggests this may create motivational challenges for students~\cite{bernstein2025beyond, budhiraja2024its}, but how AI availability shapes students' motivation to engage with programming remains an open question.

We can imagine students responding to easy access to AI tools in different ways. Some might feel discouraged, questioning why they should put effort into assignments when peers can get good grades using AI. Others may question the value of building programming skills when AI can produce working code effortlessly, a concern that extends to both current effort and future career relevance~\cite{farooqi2026job, chan2023students}. Others might find that AI tools help them write or debug code more effectively~\cite{liffiton2023codehelp, denny2024desirable}. Still others may think very little about AI. These varied reactions~\cite{bernstein2025beyond} highlight that AI availability may shape motivation in ways that extend beyond whether students choose to use it, and may reveal existing motivations that were less apparent before a faster alternative was available.

Understanding how students experience this situation matters for three reasons. First, motivation is a well-established predictor of persistence and learning in computing education~\cite{luxtonreilly2018introductory, eccles2020expectancy, gladstone2022situated}. If AI availability threatens students' motivation to engage deeply with programming, the consequences may extend beyond any single assignment to students' long-term learning and retention. Second, when motivation is low and AI provides an easy alternative, students may not do the work at all. Recent research suggests this is already happening. Stone~\cite{stone2025generative} found that 41\% of surveyed students reported using AI in ways explicitly banned, and other studies have documented students disabling pedagogical guardrails to access direct solutions~\cite{kapoor2025exploring} and shifts in academic integrity behaviors in the presence of generative AI~\cite{chen2024plagiarism}. Third, instructors need guidance navigating this landscape. Instructors currently face difficult choices about whether to ban AI entirely and risk students using it anyway, or allow it and wonder whether students are actually learning~\cite{lau2023ban}. Making informed decisions requires understanding how students experience AI availability and what helps them stay engaged with the material.

A growing body of research has explored the impact of generative AI on computing education~\cite{becker2023programming, prather2023robots, prather2025beyond, bouvier2025rest}. Much of this work has focused on what motivates students to use AI tools~\cite{margulieux2024self, amoozadeh2024trust, prather2023its}, and a few studies have begun to examine the relationship between motivation and LLM use in programming~\cite{boguslawski2024programming}. However, less attention has been paid to how the constant awareness that AI can complete their work relates to students' motivation to put effort into assignments themselves, regardless of whether they actually use these tools. We draw on Situated Expectancy-Value Theory (SEVT), which frames motivation as shaped by both individual beliefs and situation factors such as available alternatives \cite{eccles2020expectancy}. Within SEVT, the perceived value and cost of an activity are not fixed but are constructed in the moment. This makes SEVT well suited for examining how the availability of AI, as a situational factor, may relate to how students experience the effort, value, and costs of their programming assignments.

In this study, we investigated how engineering students described their motivation to put effort into their programming assignments in the context of AI availability. We focused on engineering majors specifically, as they are end-user programmers~\cite{ko2011state} for whom programming serves disciplinary engineering work rather than being the core professional skill they are training to develop. Prior work has argued that this distinction matters for motivation as non-majors form a distinct audience for introductory computing whose motivational goals should not be inferred from CS-major samples~\cite{forte2005motivation, hur2024profiling}. We conducted semi-structured interviews with 13 engineering students in a required introductory MATLAB course where students could use a sanctioned, class-specific AI tool. Our analysis, guided by SEVT, examined how students described their motivation in the context of AI availability and whether their behavior aligned with their stated values around learning. We address the following research questions:

\begin{itemize}
    \item [\textbf{RQ1:}]How do engineering students describe how AI availability shapes their motivation to put effort into programming assignments?
    \item [\textbf{RQ2:}] How do students navigate the tension between expressed value for learning through effort and the constant availability of AI as an alternative to effort?
\end{itemize}

\section{Related Work}

\subsection{AI in Computing Education}
Generative AI tools can now solve most introductory programming problems~\cite{finnie2022robots, denny2023conversing, denny2024computing}, which means for most assignments students encounter, a working solution is available on demand. A growing body of computing education research has begun to examine what this shift means for teaching and learning~\cite{becker2023programming, prather2023robots, denny2024computing}. Much of the response so far has been instructional, such as exploring tools designed to scaffold student problem-solving without providing direct answers~\cite{liffiton2023codehelp, denny2024desirable} and AI-powered instructional tools for large-enrollment courses~\cite{liu2024teaching, tabarsi2025merryquery}. While these works advances understanding of how to teach with or alongside AI, it primarily frames AI as an instructional-design problem. Our study focuses on a different question: when AI can do the assignment, how does that change students' motivation to do it themselves?

For novice programmers specifically, a growing body of research has explored the effects of AI tool use on student learning. AI tools can reduce frustration during programming tasks~\cite{kazemitabaar2023studying}, provide immediate feedback~\cite{becker2023programming}, and help students debug their code more efficiently~\cite{denny2024desirable}. However, recent work has revealed more mixed results. Learning benefits from AI were observed primarily among students with prior programming experience~\cite{kazemitabaar2023studying}, while less experienced students encountered metacognitive difficulties such as an illusion of competence~\cite{prather2024widening}. In one controlled experiment, students who lost AI access after using it performed worse than those who never had it, unless the tools were designed to scaffold rather than complete~\cite{bastani2024generative}. Margulieux et al.~\cite{margulieux2024self} found AI use patterns vary as students with higher self-efficacy and prior grades tended to use AI less or later in problem-solving. Collectively, this body of work, including recent systematic reviews~\cite{bernstein2025beyond, choudhuri2025insights}, establishes that AI's effects on student learning are complex, but it focuses primarily on learning outcomes and behavior. With respect to motivation, Padiyath et al.~\cite{padiyath2024insights} found that students' self-perceived over-reliance on LLMs was negatively associated with self-efficacy in an undergraduate programming course and that perceived peer use normalized adoption even among initially skeptical students. However, further research is needed to understand how AI availability shapes students' \textit{motivation} to invest effort in programming, whether through the experience of using AI or through awareness that AI can complete the work they are being asked to do.

\subsection{Theoretical Framework: Situated Expectancy-Value Theory}

This study uses Situated Expectancy-Value Theory (SEVT)~\cite{eccles1983expectancies, wigfield2000expectancy, eccles2020expectancy} as its analytical framework for understanding how AI availability enters into students' motivational reasoning about programming work. We chose SEVT for this investigation for two reasons. First, it decomposes subjective task value into distinct components: intrinsic value, attainment value, utility value, and cost, each of which captures a different reason a student might find a task worth pursuing or avoiding~\cite{eccles2020expectancy}. Together with expectancy, students' beliefs about their ability to succeed, these components provide a multi-dimensional account of motivation that allows us to analyze not just \textit{whether} AI availability is associated with motivational changes but \textit{which specific dimensions} students describe as most affected. Second, SEVT is a \textit{situated} theory: expectancy and value beliefs are not stable traits but are actively shaped by the task context, including the tools and alternatives available to students~\cite{eccles2020expectancy}.

SEVT distinguishes four components of subjective task value. \textit{Intrinsic value} refers to the enjoyment or satisfaction derived from doing the task itself. \textit{Attainment value} captures the importance of the task to one's identity and self-concept. Doing well matters because of who the student understands themselves to be. \textit{Utility value} reflects the perceived usefulness of the task for current and future goals, such as career preparation or course grades. Finally, \textit{cost} represents what is given up or suffered by engaging in the task, including the effort required and the alternatives forgone~\cite{eccles2020expectancy}. Unlike the other subjective values, cost works \textit{against} motivation. This multidimensionality is critical for understanding motivation in the context of AI availability, because AI does not uniformly raise or lower students' motivation to do programming assignments. A student might enjoy the challenge of programming but question whether the skill is worth acquiring when AI can produce the code. Another might recognize programming's career relevance but find the effort cost of independent work prohibitive when a faster path exists. A unidimensional account of motivation would flatten these distinctions.

Among these components, cost has received the least theoretical and empirical attention, despite its practical significance for understanding student disengagement~\cite{barron2015expectancy}. Flake et al.~\cite{flake2015measuring} advanced the treatment of cost by empirically distinguishing its sub-components: \textit{effort cost} (the perceived effort required), \textit{opportunity cost} (what else could be done with that time and energy), and \textit{loss of valued alternatives} (the persistent emotional pull of a forgone option). This distinction is relevant to the present study, where AI availability introduces a salient alternative that students must weigh against continued independent effort on programming assignments.

Because SEVT is a situated theory, it predicts that changes in the task context should reshape students' motivational judgments~\cite{eccles2020expectancy}. AI availability represents such a change. Both using AI directly (which may alter how students experience the process of programming) and simply knowing AI is available (which may reshape how students weigh the cost and value of independent effort) could enter into students' motivational reasoning about programming assignments.

\subsection{SEVT in Computing Education} SEVT and related expectancy-value frameworks have been applied in computing education to explain student engagement and persistence. Luxton-Reilly et al.~\cite{luxtonreilly2018introductory} identified motivational constructs, including perceived value and expectancy, as important predictors of introductory programming outcomes. Lewis et al.~\cite{lewis2019alignment} found that alignment between students' personal goals and their perceptions of the computing field predicted sense of belonging, connecting to SEVT's utility and attainment value dimensions. Lishinski et al.~\cite{lishinski2016learning} examined how expectancy-value beliefs interacted with gender to predict CS1 performance, and Sibia et al.~\cite{sibia2024examining} found that both perceived challenges and perceived potential played a role in students' intentions to major in computer science.  More recently, expectancy-value constructs have been connected to AI contexts. Chan and Zhou~\cite{chan2023deconstructing} applied an EVT-based instrument to student perceptions of generative AI, and Kamberovic et al.~\cite{kamberovic2025investigating} found that expectancy and value beliefs predicted AI tool usage frequency. Collectively, this work demonstrates that SEVT's constructs are relevant to computing education and that motivational beliefs shape student outcomes. However, prior applications have focused on higher-level questions, such as persistence in computing or choice of major~\cite{lewis2019alignment, sibia2024examining}, rather than how students reason about effort on specific assignments. Moreover, no prior work has applied SEVT to examine how the availability of AI tools enters into students' motivational judgments about programming work.

\section{Methodology}
\label{sec:methodology}


\subsection{Population and Study Context}

We conducted this study during the last three weeks of Fall 2025 at a large public research university. We recruited participants from an introductory MATLAB programming course designed for engineering majors. The course enrolled approximately 150 students across three sections and was structured around weekly lectures, lab sessions, and out-of-class programming assignments.

We focus on engineering majors, rather than CS majors, because being a programmer is often not their professional goal or identity. These students are end-user programmers \cite{ko2011state} who use programming as a tool in service of another discipline, rather than as a skill they are training to do professionally. Prior work has argued that non-majors form a distinct audience for introductory computing whose motivations should not be inferred from CS-major samples \cite{forte2005motivation, hur2024profiling}. This distinction matters for studying motivation because their reasons for investing effort in the course may differ from those of CS majors, whose goal is building a core professional competency.

The course operated under a six-level AI usage policy inspired by the AI Assessment Scale \cite{perkins2024artificial} that specified permissible interactions ranging from no AI use (Level 1) through increasingly collaborative roles (Levels 2--5), such as AI as thought partner, reviewer, analyst, and co-creator, to prohibiting direct solution generation (Level 6). Each level carried corresponding disclosure requirements, with higher levels requiring students to document AI contributions explicitly. The instructor also provided access to a course-specific AI assistant built on retrieval-augmented generation of course materials while allowing the instructor to monitor student interactions \cite{tabarsi2025merryquery}. We chose this context because AI was neither prohibited nor unrestricted and students navigated multiple tools and usage levels with explicit agency over their choices.
\begin{table*}[h]
\caption{Participant Demographics}
\label{tab:demographics}
\begin{tabular}{llllll}
\toprule
\textbf{ID} & \textbf{Gender} & \textbf{Year} & \textbf{Major} & \textbf{First-Gen} & \textbf{Prior Prog. Exp.} \\
\midrule
P1  & Male   & Freshman  & Mechanical Eng.  & No  & Some \\
P2  & Male   & Freshman  & Aerospace Eng.   & No  & None \\
P3  & Male   & Freshman  & Electrical Eng.  & No  & Some \\
P4  & Female & Sophomore & Biomedical Eng.  & Yes & None \\
P5  & Male   & Freshman  & Mechanical Eng.  & No  & Some \\
P6  & Male   & Freshman  & Computer Eng.    & No  & None \\
P7  & Female & Freshman  & Civil Eng.       & Yes & Some \\
P8  & Male   & Sophomore & Aerospace Eng.   & No  & None \\
P9  & Male   & Junior    & Mechanical Eng.  & No  & Extensive \\
P10 & Female & Freshman  & Electrical Eng.  & Yes & None \\
P11 & Male   & Freshman  & Aerospace Eng.   & No  & None \\
P12 & Male   & Sophomore & Civil Eng.       & No  & Extensive \\
P13 & Female & Freshman  & Mechanical Eng.  & No  & None \\
\bottomrule
\end{tabular}
\end{table*}

\subsection{Participants}
\label{sec:participants}

After receiving IRB approval and course instructor permission, we recruited participants through email announcements, course forum posts, and an in-person classroom visit. Interested students completed a consent survey and all participants consented to audio recording.

To focus on students actively navigating AI availability in their coursework, we recruited during the final weeks of the semester when students had accumulated substantial experience with both programming assignments and AI tools. In order to limit selection bias, and to attract students who may not have otherwise been interested in the topic, the instructor agreed to offer a small amount of extra credit toward a lab assignment to students who completed the interview. We acknowledge that self-selection may have biased our sample towards students with stronger opinions about AI. We discuss this limitation further in Section~\ref{sec:limitations}. The first author had no role in the course and conducted interviews outside of class time, reducing perceived pressure on participants to align responses with course expectations.

Of the students who expressed interest, 13 completed interviews. Participants were predominantly White (n=9), with three Asian students and one Black or African American student. Three participants were first-generation college students. Table~\ref{tab:demographics} provides individual participant details.

We categorized participants' prior programming experience based on how they described their backgrounds during interviews, rather than through a predetermined survey measure. We chose this approach because participants' open-ended descriptions captured meaningful distinctions that a single survey item would not, such as the difference between having taken a high school course and actively building personal programming projects across multiple languages. Based on these self-descriptions, participants fell into three groups: no prior programming experience (n=7), some prior experience from high school or other contexts (n=4), and extensive programming experience with personal projects (n=2). Most participants (10 of 13) reported using outside AI tools such as ChatGPT for their programming work, while seven reported also using the course-provided AI assistant.

\subsection{Semi-Structured Interview}
\label{sec:interview}
Each semi-structured interview lasted approximately 30 minutes and was conducted via Zoom by the first author. The protocol consisted of three parts.

\textbf{Programming Context}: We began each interview by asking participants about their coursework and programming experience. For example, we asked them to describe recent programming assignments, identify core skills they were learning, and describe any prior experience using AI in their programming process. These questions established shared context about participants' course experiences and initial AI exposure before the structured motivational questions that followed.

\textbf{AI Capability Demonstration}: To establish shared understanding of what AI tools can accomplish, participants viewed a brief video demonstration\footnote{\url{https://www.youtube.com/live/0Uu_VJeVVfo?t=1088s}} of approximately 60 seconds illustrating generative AI's current programming capabilities, including code generation from natural language descriptions. We included this demonstration because we recognized that participants' understanding of AI capabilities might vary substantially, and we wanted to ground the subsequent conversation in a concrete shared reference point and establish shared context before eliciting reflections. 

\textbf{Motivation and Experience Questions}: The bulk of each interview focused on how AI availability intersected with students' motivation to put effort into assignments. We deliberately framed questions around \textit{effort}, rather than completion or AI usage, because we were interested in participants' willingness to try and persist, which more directly captures motivation. We drew from SEVT \cite{eccles2020expectancy} to structure these questions around distinct motivational dimensions. For each dimension, we first asked participants to describe their experience in general terms, then asked specifically whether and how AI had played a role. For example:

    \begin{itemize}
        \item ``How enjoyable do you find it when you put effort into a programming assignment for class?'' (intrinsic value)
        \item ``In what ways, if any, has AI affected how enjoyable you find it?''
        \item ``How confident do you feel that putting effort into completing programming assignments will help you succeed?'' (expectancy)
        \item ``If at all, how has the availability of AI tools influenced that confidence?''
    \end{itemize}

We repeated this pattern across all five SEVT dimensions (intrinsic value, utility value, attainment value, expectancy, and cost). We also asked participants about their perceptions of peers' AI use to capture social dynamics.

\subsection{Data Analysis}
\label{sec:analysis}

\textbf{Positionality}: The first author, a fourth-year doctoral student with prior qualitative research experience studying how students program with AI, conducted all interviews and led the analysis. The second author, a postdoctoral researcher with qualitative methods expertise, served as a collaborative analytical partner throughout. The third author's research focuses on AI- and data-driven learning environments that support novice programmers, with growing interest in how generative AI affects how those students engage with programming. He approached this work in an advisory role, supporting the framing and analytical direction without coding transcripts directly. We note our positionality because reflexive thematic analysis treats the researcher's perspective as a resource rather than a source of bias \cite{braun2019reflecting, braun2024rtarg}. The first author is broadly interested in AI's potential to support programming learners but entered this study more aware to possible negative effects on motivation, and was struck by how much these early-career engineering students, who were not CS majors, had already processed AI's role in their learning. Listening to how students talked through their AI use drew our analytical attention to the contradiction between what they said they valued and what they described doing with AI.

\textbf{Analytical approach}: We analyzed our interview data using reflexive thematic analysis \cite{braun2019reflecting, braun2024rtarg}, an interpretivist approach in which the researcher actively constructs themes through sustained engagement with the data, rather than discovering pre-existing patterns within it \cite{padiyath2026reflecting}. We chose this approach because our research questions asked how students made meaning of their motivational experiences, a goal better served by interpretive construction of themes than by coding reliability procedures. Consistent with reflexive thematic analysis, we did not calculate inter-rater reliability metrics, as such measures reflect a positivist logic of coding accuracy that is incoherent with the interpretivist values underlying our approach \cite{braun2021onesize}.

\textbf{Coding process}: Our analytical process began with the first and second authors collaboratively coding one interview transcript to develop an initial set of codes and a shared understanding of how to approach the data. We then independently coded two additional transcripts and met to discuss differences, not to reach consensus, but to deepen our interpretations and refine code definitions through critical dialogue \cite{braun2024rtarg}. Our coding combined deductive and inductive orientations. The deductive orientation used SEVT constructs to organize codes for responses to dimension-specific questions, mirroring how our interview protocol was structured. The inductive orientation captured patterns participants raised outside the SEVT framework, such as peer comparison, willpower, and resistance to AI use. After establishing this shared analytical approach, the first author served as the primary coder for all 13 transcripts, with the second author reviewing coded transcripts and flagging disagreements for discussion. Because our coding was organized around SEVT dimensions, the thematic structure of our findings followed the same framework. Within each dimension, we examined how participants described their motivation and how AI availability intersected with it, reviewing patterns across the full dataset through discussion between the two authors. 

\textbf{Emergent pattern}: Through this process, we identified a pattern that cut across dimensions: participants described valuing effort and struggle in their programming coursework, yet described using AI to complete their assignments. This pattern captured the phenomenon participants described more fully than any single SEVT dimension, and inspired our second research question (RQ2), which emerged from the data rather than being specified a priori.

\section{Findings}

In this section, we first describe students' baseline motivation to put effort into programming assignments \textit{in general}, when asked without referencing AI (Section~\ref{sec:baseline}). We then discuss how their awareness of AI's capabilities, regardless of their own use, shaped their motivation across SEVT dimensions (Section~\ref{sec:effect}). Finally, we describe how students navigated the tension between their stated values around effort and struggle and the easier path presented by AI (Section~\ref{sec:tension}). Where helpful, we report participant counts (out of 13) to give readers a sense of each pattern's scope, though our qualitative sample does not support claims about prevalence.

\subsection{Baseline Motivation to Put Effort into Programming Assignments}
\label{sec:baseline}
Our interview protocol first asked students about their motivation to put effort into programming assignments without specifically referencing AI (see Section~\ref{sec:methodology}), though because the protocol alternated between general and AI-specific questions for each dimension, AI likely became more salient as the interview progressed. Students did sometimes bring up AI in their answers. This section describes what students reported across expectancy (beliefs about their ability to succeed) and four subjective task values: intrinsic value (enjoyment derived from effort), attainment value (importance to their identity), utility value (usefulness for their goals), and cost (effort and competing demands).

\subsubsection{Expectancy: Belief About Ability}


Most students expressed confidence in their ability to complete programming assignments through their own effort. P12 described themselves as someone who picks ``\textit{up on stuff really quickly,}'' P7 reported feeling ``\textit{very confident in my abilities to finish an assignment,}'' and P4, despite self-identifying as ``\textit{not a coder really,}'' stated: ``\textit{it never ever feels like something I'm not capable of doing.}'' Students who expressed lower confidence did so in relation to specific tasks rather than general ability. P6 acknowledged that debugging independently would be difficult, and P8 described being ``\textit{still fairly new}'' to coding. No student described themselves as unable to complete their assignments.

Not all students could cleanly separate their confidence from AI availability. When asked how confident they felt that putting effort into their programming assignments would yield positive results, P10 immediately asked: ``\textit{Like, with AI or without AI?}'' Though prior interview questions may have increased AI's salience, P10's need for clarification suggests that for some students, confidence was difficult to discuss independently of AI availability.

\subsubsection{Intrinsic Value: Enjoyment of the Process}

Our interview questions asked specifically about enjoyment of putting effort into programming assignments, so the patterns described here reflect that framing. Most students described having some level of enjoyment from this effort, with five (P2, P5, P7, P8, P12) expressing strong motivation tied to the struggle itself. For example, P5 said: ``\textit{I like trial and error, so when I put in effort into something and it works, it's always so fun. I like that part of coding a lot.}'' P12 described the relationship between struggle and reward: ``\textit{Whenever you struggle for a really, really long time, and then finally succeed, it's super gratifying.}'' P2, a first-time programmer, recalled a debugging problem whose solution ``\textit{came to me in a dream}'' as ``\textit{probably the most excited I had been in a course this whole entire semester.}'' P8, also relatively new to programming, described learning to code as: ``\textit{It's kind of like making art, or doing a sport. It takes practice, and then boom, the results are there.}'' For these students, programming was enjoyable because it required effort, and the struggle made the outcome feel earned.

Enjoyment was not universal, and for some students it was conditional. P6 described enjoyment that depended on whether breakthroughs occurred: ``\textit{If I really can't figure it out and I'm just stuck, it's not very enjoyable. But once I've powered through and I've figured out how to do something, it does make me feel good about myself.}'' For P6, enjoyment was contingent on eventually solving the problem rather than on the effort itself. Other students found the effort ``\textit{very enjoyable}'' (P7) and ``\textit{exciting}'' (P4), though with less elaboration on why.

Not all students shared this sentiment. P1 stated plainly: ``\textit{I can't say I'm a big fan of coding,}'' finding longer projects with ``\textit{200, 300 lines of code}'' to be ``\textit{a little tedious and kind of not enjoyable,}'' though P1 appreciated the work when it ``\textit{can prove useful for the user.}'' P3 similarly noted: ``\textit{I will say I do not enjoy programming.}'' P9 presented a different pattern: high intrinsic value for personal projects (``\textit{Java game development, just for fun}'') but low enjoyment for coursework, which P9 completed just to ``\textit{get the grade.}''

\subsubsection{Attainment Value: Identity and Ethics}
When asked about personal values and effort, some students connected programming effort to their self-concept, while others focused on pragmatic concerns. P3 stated most directly: ``\textit{I picked this major for the challenge. I love the challenge. That's who I am.}'' P3 added: ``\textit{The point is to struggle. The point is to fail. The point is to learn. That's how growth occurs.}'' Other students expressed strong values around independent effort without using explicit identity language. P5 stated: ``\textit{If I can do it myself, I will do it myself, because I feel I learn better if I'm doing it.}'' P2, who described themselves as ``\textit{kind of stubborn}'' about working independently, framed understanding as a matter of reputation: ``\textit{if you get asked a question and you've used AI your entire class and you don't know what they're talking about, it's kind of not a really good reflection of you.}''

A few students named integrity as an important value (P1, P3, P12). P1 stated: ``\textit{I think the ethical thing to do is to learn the material and try your best... ultimately it's important for your integrity to do assignments honestly.}'' P3 used strong language: ``\textit{I think it shows a moral failure if you're not able to make yourself do the work.}'' For these students, effort reflected values about how one should approach learning, not just what outcomes it would produce. Not all students framed effort in these terms, with several (P1, P6, P9, P10, P11, P13) focusing on grades or pragmatic concerns rather than self-concept.

\subsubsection{Utility Value: Career and Grades}

Students valued programming effort at different timescales. For some, the primary utility was grades (short-term). P6 admitted: ``\textit{I don't really care, but I care in the fact that I want my GPA to stay good.}'' P10 was similarly grade-driven: ``\textit{I'm trying to get an A.}'' For others, they saw programming as essential for their engineering careers (long-term). P8 stated that programming would be ``\textit{very, very beneficial}'' for handling complex calculations in their future job, P13 viewed MATLAB as ``\textit{an industry standard}'' and found it ``\textit{pretty important to me}'' to know deeply rather than rely on AI. P1 argued that ``\textit{knowing MATLAB, knowing C, knowing Python is going to be something that's going to differentiate you.}'' P3, despite finding no intrinsic enjoyment, saw programming as essential preparation: ``\textit{the objective is to learn the material, so that when I graduate from this place, I'm able to do it without outside assistance.}'' Not all students shared this view. P11 described programming as ``\textit{something that won't be as important in the future,}'' rating it ``\textit{a little bit less important than my other classes like calculus or physics.}''

\subsubsection{Cost: Effort and Competing Demands}

Students also noted the \textit{costs} of putting effort into programming assignments. The majority of participants (n=7) were novices taking their first programming course, and many (n=7) described the work as demanding. P4 recalled the difficulty of their recent programming assignment: ``\textit{It was difficult, for sure, trying to figure out how to get all the functions to work and get everything to work at the right time.}'' P2 described the time cost of debugging: ``\textit{spending so much time trying to get things figured out and trying to figure out why one line of code isn't working for, like, 30 minutes.}'' Beyond the time investment, several students (n=5) described an emotional toll from the repeated cycle of errors and failed attempts. P8 captured this frustration: ``\textit{It definitely kind of sucks when there's a lot of errors... I can't see it. And I think that it does kind of create a frustration, and it does take hours.}'' P5 described the experience more bluntly as ``\textit{learning how to get punched in the face repeatedly by your computer until you figure it out.}'' Students also described \textit{opportunity costs}. P11, a mechanical engineering major, noted that coding ``\textit{doesn't seem to be a huge part of what I'm doing}'' and placed it below courses like ``\textit{calculus or physics}'' in priority. P10, a first-time programmer, described one project taking ``\textit{3 or 4 days to complete,}'' which ``\textit{made me spend time on that, rather than studying for other classes.}'' For these students, time spent on programming came at the expense of courses they viewed as more directly relevant to their majors.

\subsection{Motivation in the Context of AI Availability}
\label{sec:effect}

Section~\ref{sec:baseline} described students' baseline expectancy and subjective task values (intrinsic, attainment, utility, and cost). In this section, we examine how students described AI availability as shaping these dimensions, both indirectly (from knowing AI's capabilities and that they \textit{could} use it) and directly (from actually using AI). We do not include attainment value in this section because students did not describe AI as changing their personal values or sense of identity. Instead, as Section~\ref{sec:tension} examines, these values served as a resource students drew on when navigating AI availability. Nearly all students in our sample, including those who valued putting effort into their assignments, described some impact from AI on their motivation.

\subsubsection{Expectancy: Confidence That Depended on AI Being Available}
\label{sec:contingent}

Five students (P6, P8, P10, P11, P13) described confidence that appeared to depend on AI being available. P10, for example, immediately referenced AI when asked about their confidence in completing programming assignments: ``\textit{AI has definitely made it easier for me. So if I was to not use AI at all, I would probably fail the class.}'' When asked directly how their confidence would change without AI tools, other students described substantial drops. P8 acknowledged: ``\textit{If they were unavailable, yeah, I would say my skills would go down to a six [from an eight].}'' P6 corroborated this: ``\textit{It would definitely be a lot harder... I'd have to debug everything myself,}'' adding that without AI, ``\textit{I would not be nearly as confident.}''


Beyond decreased confidence, some students described uncertainty about the boundaries of their own knowledge. P10 captured this tension: ``\textit{Do I know this, or is it just AI doing it for me?}'' P11 noted: ``\textit{If I'm not doing the critical thinking on my own, then it's gotta make you a little bit worried about the tests.}'' P13 described a broader concern: ``\textit{I feel like I'm not as proficient as maybe someone pre-AI. I feel like I'm not getting as much out of an introductory course as I should.}'' P10 illustrated a particularly telling tension. When asked about working without AI, P10 acknowledged that doing so would make them ``\textit{more inclined to succeed}'' because they would be ``\textit{basing it off my own knowledge.}'' Yet in the same interview, P10 stated they would ``\textit{probably fail}'' without AI. P10 recognized that independent work would build more genuine competence but described feeling too dependent on AI to pursue it.  P10 reported studying ``\textit{more than I probably need to, just because I need to make sure I know everything, and it's just not AI.}'' The pattern P10 described suggests a potential feedback loop where AI use may erode competence for independent work, which in turn makes further AI use feel necessary.

Not all students who recognized this risk experienced it. P7 described confidence that depended on how AI was used, ``\textit{If I put in the work, and I only use AI rarely to understand something better, then I'll be able to yield good results. Versus if I'm using AI constantly, I won't learn as much, and I'll forget it faster or something. And then I would lose confidence.}'' P7 was aware that heavy AI use could erode confidence but described calibrating usage to prevent that outcome. Where P10 described being caught in the feedback loop, P7 described having avoided it through bounded use. In contrast, other students' confidence was unaffected or even bolstered by AI's limitations. P3 argued that ``\textit{if you know nothing about what you're doing, the AI knows equally nothing or just as little,}'' warning that AI output ``\textit{can be a trap for many students who just copy what it says.}'' P2, who limited AI use, stated: ``\textit{I think I would still feel pretty confident. I feel like I would still perform pretty well.}''

\subsubsection{Intrinsic Value: Reduced Enjoyment When AI Precludes Struggle}

Most students who used AI (8 of 13) reported feeling less accomplished, and those who relied on it more heavily described a greater loss of satisfaction.

This pattern was widespread among students who used AI. P10 described it directly: ``\textit{I definitely feel less accomplished after I use AI, just because I didn't do it myself.}'' P13 described a similar loss: ``\textit{Whenever I use AI, it just takes away that feeling of, `oh, I did this on my own.' It's just, `well, I guess I got it done.'}'' P7, who used AI selectively, noted that AI-assisted work ``\textit{felt more artificial}'' and that the team ``\textit{felt less accomplished than if we just worked a little bit harder on it.}'' P8 acknowledged ``\textit{losing that feeling of accomplishment a tad bit, because you're using help.}'' P11 similarly noted that AI ``\textit{lessens}'' how accomplished the work feels ``\textit{by a little bit.}'' Students described AI availability as diminishing the enjoyment that came from working through challenges independently.

However, the relationship between AI use and sense of accomplishment may not be binary. P7 described a threshold: ``\textit{We'll feel less accomplished if we just use AI versus if we use it a little bit, then we'll still feel the same amount of accomplishment.}'' This suggests that bounded use could preserve some intrinsic reward, while heavy reliance can undercut it. Though we cannot confirm a causal relationship from these accounts, the pattern was consistent across usage levels. Even P2, who limited AI use to lab-time troubleshooting, acknowledged that ``\textit{it takes away my sense of accomplishment, because I'm a pretty stubborn person, so I like to figure things out myself.}''  Students who used AI for code generation (P10, P13) described a more persistent sense of loss, while those who limited use to explanation or debugging (P2, P7) described the reduction as smaller but still present.

P9 was an exception, reporting that AI ``\textit{made it more enjoyable actually for [personal] projects because there's a lot of working with the AI,}'' though this enjoyment was tied to personal projects rather than coursework, where P9 had low intrinsic value to begin with. P6 and P10 described a similar effect: P6 found that AI made debugging less tedious, and P10 described the work as more ``\textit{enjoyable because it's easier, honestly. I know what I'm doing wrong, and it's pretty straightforward, and how to fix it. And as I'm fixing the code while using AI, I'm learning through that.}''

\subsubsection{Utility Value: Questioning the Worth of Programming Effort}

In Section~\ref{sec:baseline}, students described the utility of programming effort at different timescales, from assignment grades to career relevance. When asked directly about AI, students questioned each of these. We organize this section by timescale, from immediate academic outcomes (short-term) to career concerns (longer-term).

\textbf{Questioning the value of homework effort}: Several students (n=6) questioned whether putting effort into programming assignments was worth the time when AI could produce similar quality artifacts faster. P6 captured this directly, ``\textit{Why would I spend an hour on an assignment when I can have ChatGPT do it for me in 10?}'' P13 similarly questioned the value of effort when peers could achieve equal or better grades through AI with less work (see Section~\ref{sec:reshaped_cost}). P12 extended this concern to effort in general: ``\textit{With people using AI all the time, it devalues the content that people are actually making... it makes everything feel like more of a waste of time.}''

\textbf{Questioning the value of programming skills}: Beyond individual programming assignments, several students (n=6) questioned whether programming skills themselves were losing value, when asked whether AI availability had influenced how they saw the value of programming for their future career. P11 acknowledged: ``\textit{If it keeps progressing how it is now, then a lot of what I did in this class might not be useful within a couple of years},'' though P11 added that the skills would ``\textit{still be worth it}'' as a near-term foundation.

P10 observed more broadly that AI ``\textit{diminishes the value of coding, just because AI can code for you.}'' P12 similarly noted: ``\textit{I feel like the fact that we have AI makes learning coding not obsolete, but a little bit less important... the coding aspect is a little bit less important than it was beforehand.}'' P7 hypothesized that: ``\textit{Knowing certain functions off the top of your head is a little bit less valuable.}'' However, several students described a shift in which skills mattered. P11 noted that while writing code felt ``\textit{a little bit less important,}''  the ``\textit{problem-solving aspect of it has probably increased some more.}'' For these students, they viewed AI as not making programming irrelevant but shifting which parts of programming remained worth learning.

Some students extended this concern beyond their own careers. P5, who mentioned rarely using AI tools, was worried about being replaced: ``\textit{AI can definitely do the fundamentals, so since I'm only learning the fundamentals... I could be easily replaceable with an AI if I was to bring this set of skills to a team in the future}.'' P11 expressed concern for peers in computer science majors specifically, ``\textit{They know what's going on now, and eventually it might not mean anything, the major that they're in now.}'' These concerns suggest that for some students, AI availability raised questions not just within the context of assignment or course but the whole trajectory of programming as a career-relevant skill.

\textbf{Programming as a valuable skill in the face of AI}: Many students (n=7) continued to see programming knowledge as essential despite AI's capabilities. P8 maintained that programming would still be ``\textit{very, very beneficial}'' for handling complex calculations, and P1 argued that programming knowledge would ``\textit{differentiate you.}'' Some students (P2, P3, P5) went further, arguing that AI made programming knowledge more valuable. P3 explained: ``\textit{I still think it's incredibly important, if not more so now than it was before, to understand the basic concepts of programming, so that in an age where AI can make programming better than it has been ever before, you are equipped with the mindset and the disciplinary knowledge to do so.}'' For these students, the relationship between AI availability and perceived career utility was not uniformly negative.



\subsubsection{Cost: AI as a Visible Alternative to Effort}
\label{sec:reshaped_cost}

\textbf{Effort feels less worthwhile}: Several students (n=6) who expressed strong baseline motivation described how knowing that AI could complete the assignment changed how worthwhile their effort felt. P4, who found coding ``\textit{exciting}'' and expressed confidence in their ability to learn, described the effect: ``\textit{I'm spending all this time doing it, and then an AI could do it in, like, 3 seconds. Like, I guess, objectively, sometimes that is kind of demoralizing.}'' P12, who described coding as ``\textit{super gratifying,}'' acknowledged: ``\textit{It's really demotivating, to put it flatly.}''  P11 pointed to a specific scenario, where knowing AI could find a bug instantly ``\textit{makes it a little bit less worthy of your time to go and spending hours upon hours trying to find a bug}'' yourself. For these students, the issue was not that they lacked motivation, but that knowing a faster alternative existed made the cost of effort more noticeable. The same hours of debugging felt different when an alternative existed that could bypass them entirely.

\textbf{A faster path to existing goals}: For other students, they mentioned that AI did not so much change their motivation, but rather provided a more efficient route to outcomes they were already focused on. P6, whose baseline motivation was primarily grade-driven (``\textit{I don't really care, but I care in the fact that I want my GPA to stay good}''), described it clearly: ``\textit{Knowing that I have AI and I can just get it done with [it] is really tempting.}'' When facing competing demands from other courses, P6 further explained, ``\textit{When I have some big assignments due for other classes, I'm not as concerned with MATLAB for that day or that week.}'' For students like P6, AI did not necessarily lower their motivation, but it offered a practical solution to an existing challenge that many students face, which is limited time spread across multiple demanding courses. P10, also primarily grade-driven (``\textit{I'm trying to get an A}''), described AI as shifting how effort was allocated: ``\textit{It makes me only want to understand the concepts when I'm getting ready for a test, rather than doing the assignments itself.}'' For these students, AI did not so much change their motivation as provide a way to manage costs they already found difficult.

\textbf{Reduced friction}: Not all cost perceptions were negative. Several students (n=5) described AI as reducing one specific cost, the frustration of being stuck. P6, when asked whether AI affected how enjoyable programming felt, responded, ``\textit{It's definitely more enjoyable, because if I don't figure something out, I can paste the code... and I can ask AI, like, what am I doing wrong?... It just makes it a lot quicker to resolve, so I can go back to figuring it out and doing it correctly.}'' P8 described a similar situation: ``\textit{in times where I'm, like, struggling, AI does take that load off. So AI does make me feel better.}'' For P6 and P8, AI reduced the emotional cost of programming, not just the time cost, by removing the frustration of being stuck.

\subsubsection{How AI Availability Reshapes Value Through Social Comparison}
The previous subsections focused on how students described their own relationship with AI. However, several students (P5, P6, P11, P12, P13) described their effort feeling diminished not by their own AI use, but by awareness that peers could or did use it. This social dimension cuts across multiple SEVT constructs. P5 explained: ``\textit{Someone else could have done this with AI, or someone tells me I could have just done that with AI... it just feels awkward in my head. Like, I just put in effort and you ask [AI] to put in the effort for you.}'' P12 put it more directly: ``\textit{With people using AI all the time, it devalues the content that people are actually making, and it doesn't feel as valuable because it's not treated special anymore.}'' P13 framed it in terms of fairness: ``\textit{there's these people that use AI, can get a higher grade than me for putting in less effort. So it's just, why put in this much effort?}'' P5 and P12 maintained their preference for working independently despite this frustration, while P13 described it as pulling them toward AI use.

Other students described a different response. Rather than frustration, peer AI use created a pull toward conformity. P11 described a threshold, ``\textit{If I thought it was 90\% of the class that is doing it, then I would definitely be more motivated to use it.}'' At an estimated 50\% usage, P11 did not feel compelled. P6 described a similar awareness, noting that many peers ``\textit{have waited, or now they feel comfortable to wait for the day of to complete the homework, because they know that if they can't figure it out ... they can just get AI to give them a response immediately.}'' Where P5 and P12 described peer AI use as undermining the value of their own effort, P11 and P6 described it as normalizing AI use and lowering the barrier to joining in.

\subsubsection{Situational Triggers for AI Use}
Regardless of which motivational pattern students described, certain situations intensified the desire to save effort with AI. When frustration from debugging exceeded a threshold, AI became the default response. P13 described this tipping point: ``\textit{If I just can't see what's wrong with my code, I just kind of slap it into chat and say, what's wrong with this?}'' P8 described a similar moment: ``\textit{I had to just use AI after the lab because I was just stuck.}'' P7, working under deadline pressure, noted: ``\textit{we were on a bit of a time crunch, so we had to use our resources.}'' These moments suggest that even students with strong baseline motivation could be drawn toward AI use when situational factors, such as frustration, time pressure, or competing deadlines, made the cost of independent effort feel too high.

\subsection{Navigating the Tension}
\label{sec:tension}

We now examine how students navigated the tension between their expressed value for learning through effort and the availability of AI as an alternative. Nearly all students (12 of 13) described valuing learning through effort while also feeling drawn toward using AI to reduce that effort. What varied was how students acted on those values. The majority (n=7) described limiting their AI use in ways consistent with their stated values. Five students described using AI more extensively, in ways they themselves viewed as undermining their learning. One student (P9) fell outside this framework, as their disengagement from coursework was independent of AI availability.



\subsubsection{Limiting AI Use to Protect Learning (n=7)}
\label{sec:managing}

Seven students described limiting their AI use in ways consistent with their learning values. These students were still drawn toward using AI, as P4 called it ``\textit{objectively demoralizing}'' and P12 called it ``\textit{really demotivating,}'' but they described strategies that helped them act on their values despite this. For some, these strategies were anchored in identity and personal values described in Section~\ref{sec:baseline}, such as P3's connection between programming effort and personal identity and P2's concern for professional reputation. What distinguished these students was not stronger values, but practical ways of maintaining them.

\textbf{Reframing What Makes Effort Worthwhile:} For students who found genuine enjoyment in struggle, AI's availability as an alternative could undermine the satisfaction that made programming worthwhile. Some students described reframing their focus in ways that may have helped preserve that enjoyment.

For example, P4 who found coding ``\textit{exciting,}'' acknowledged the tension, ``\textit{Sometimes I'm like, wow, what's the point of this?}'' but actively reframed the situation ``\textit{I kind of try to shift my mindset to more like, it's about the learning process}.'' P4 went on to explain what AI cannot replicate, ``\textit{learning how to deal with the challenges, learning how to troubleshoot... that is not something that AI can really help you with}.'' P5 and P12, who described coding as ``\textit{fun}'' and ``\textit{super gratifying}'' focused on the craft itself. P5 put it simply, ``\textit{There's nothing wrong with AI doing the coding, it's just, I want to code too!}''  P12 compared their work with their peers using AI, ``\textit{my assignments were better because I was putting more soul into them.}'' For these students, AI did not take away their intrinsic motivation.



\textbf{Setting boundaries on AI use}: Bounding AI use was the most common practical strategy that made acting on their values easier. Rather than avoid AI entirely, students drew lines that preserved learning. For example, P4 restricted use to syntax checking because the line between help and answer felt ``\textit{very thin.}'' P3 mentioned only using AI ``\textit{to check my work after giving it my best.}'' P12 suggested a litmus test: ``\textit{after you use AI, I feel you should be able to recreate the same code or essay or math without. I feel if you can no longer do that, then that's when you cross the line.}''


\subsubsection{When AI Use Conflicted with Stated Values (n=5)}
\label{sec:dissonance}

Five students described using AI more extensively, in ways they themselves viewed as conflicting with their stated values around learning and effort.

\textbf{Stated values and conflicting behavior:} These students were not lacking in stated motivation. P10 explained that independent work was meaningful: ``\textit{whenever I do it myself, I feel pretty good and I feel like, okay, I actually got it,}'' emphasizing that ``\textit{understanding where it's coming from and why it's working is a big factor for me.}'' Similarly, P13 acknowledged: ``\textit{I know if I don't use AI... I'm actually learning a lot more.}'' Yet these same students described AI use that conflicted with those values. P10 described routinely screenshotting errors and pasting them into ChatGPT, which ``\textit{just outputs, like, a brand new code that I use.}'' P6 described bypassing independent debugging: ``\textit{instead of having to go to office hours... it was just a quick copy and paste, quick question, and basically within a minute, we'd have an answer.}'' P13 described a similar pattern: ``\textit{sometimes I do get a little lazy with debugging, so if something doesn't immediately work, I just paste it into AI.}''

\textbf{What distinguished these students:} Where students who limited their AI use drew explicit boundaries, these students described more permissive interpretations of the course's AI policy. P6 noted that policies were ``\textit{a gray area on how much you're allowed to use.}'' P10 interpreted the course's AI permission broadly: ``\textit{[the instructor] is allowing us to use AI, so I might as well just use it.}'' Rather than drawing their own lines, these students described their situation through fairness comparisons. As P13 put it: ``\textit{there's these people that use AI, can get a higher grade than me for putting in less effort. So it's just, why put in this much effort?}'' These students also described learning as important to them, but unlike students such as P3 who connected programming effort to their identity, they did not describe it as part of who they were.

Prior programming experience may also have played a role. Students who limited their AI use included several with prior programming experience (P3, P5, P12), while students who used AI more extensively were largely encountering programming for the first time. P11, a first-time programmer, observed that less experienced students ``\textit{definitely lean on [AI] more than the people that already know a lot about coding, because sometimes whenever you just don't want to get over that hump of actually learning, then it's really enticing to just do it really quick.}'' Students with prior experience may have had less need for AI and therefore less opportunity for their behavior to diverge from their stated values.

\textbf{Using AI in targeted ways:} Two students (P8, P11) described using AI in more limited ways that appeared to reduce the tension. P8 used AI primarily as a learning tool, acknowledging ``\textit{losing that feeling of accomplishment a tad bit}'' but emphasizing that ``\textit{it does help me understand more about what went wrong.}'' P11 described limiting use to ``\textit{small things}'' but noted a persistent unease: ``\textit{whenever you use it, you always have this sense that it's not really what you're supposed to be doing.}''


\textbf{Consequences}: The three students who used AI most extensively (P6, P10, P13) described negative consequences they could name, including reduced accomplishment and uncertainty about their own knowledge. P13 also described observable performance impacts: ``\textit{There have been a few times where I have relied heavily on AI with certain modules and stuff. And I noticed I do pretty poorly in those modules I rely heavily on AI for, versus the ones where I'm doing it myself.}'' P10 captured the overall trade-off: ``\textit{I gain more knowledge, but less skill. It tells me what to do and what codes mean, but just because I'm not doing it myself, it kind of retracts from the point of doing the assignment.}''

\section{Discussion}

\subsection{RQ1: AI Availability and Student Motivation}

Our findings suggest that AI availability can do more than provide an alternative to effort; it may actively shape engineering students' motivation to put effort into assignments. The students in our study who described reduced learning and growing dependency on AI still expressed genuine value for learning and interest in the material. The issue for many participants was not a lack of motivation but a shift in how worthwhile their effort felt when a faster alternative was always available. Prior work has documented that AI use can widen learning gaps between students~\cite{prather2024widening} and that AI benefits were found disproportionately among students with stronger CS foundations~\cite{kazemitabaar2023studying}, but has not examined the motivational mechanisms through which AI availability may shape how students reason about the value of their effort. 

Students described both direct and indirect effects. The direct effects came from AI itself, such as a reduced sense of accomplishment and diminished enjoyment when AI shortened the problem-solving process. The indirect effects came from simply knowing AI was available, such as questioning whether programming skills would remain valuable and feeling that their effort was devalued when peers could achieve similar outcomes with less work. Not all effects were negative. Two students described AI as reducing frustration, consistent with findings from prior work~\cite{kazemitabaar2023studying}. Many students in our sample placed value, and even personal identity, on learning through effort. Similar patterns appear in prior computing work: Price et al.~\cite{price2017factors} found that nearly all novice programmers in their study expressed a desire to ``figure it out myself,'' and Marwan et al.~\cite{marwan2020unproductive} documented that students frequently avoid help even when it would benefit their learning. This finding may also reflect selection bias in our sample, as students who volunteered for interviews about programming motivation may have been more motivated than average (see Section~\ref{sec:limitations}). 

SEVT provides a useful framework for unpacking these effects~\cite{eccles2020expectancy}. Because SEVT treats motivation dimensions as independent, students could simultaneously value learning (intrinsic and attainment value) while finding the effort to sustain independent work increasingly difficult to justify (cost). Our results reported effects across the SEVT dimensions (Sections 4.2.1-4.2.4), but cost was the dimension students described most consistently~\cite{flake2015measuring}. Understanding what drives the cost of resisting AI requires looking at how students interact with help more broadly. The help-seeking literature distinguishes instrumental help-seeking, where students seek only enough assistance to solve problems themselves, from executive help-seeking, where students seek to have someone else complete the task on their behalf~\cite{nelson1986help}. Prior work in programming has documented both patterns, with students abusing hints by clicking through to answers without engaging with the content~\cite{marwan2020unproductive}, and avoiding help entirely out of a desire for independence~\cite{price2017factors}. AI may introduce a qualitatively different help-seeking context. While hints tend to be instructor-designed, scoped to specific steps, and delivered at discrete moments, AI is permanently visible and capable of completing the entire task, making the executive path far more accessible than it has been with prior forms of help. At the same time, this challenge is not entirely new. Students have long struggled to use help instrumentally rather than expediently, and AI availability may intensify rather than create it. 

\subsection{RQ2: Navigating the Learning-Expedience Tension}

Our results revealed differences in how engineering students navigated the tension between their learning goals and the incentives to use AI to complete work expediently (Section 4.3). Prather et al.~\cite{prather2024widening} documented a widening gap between students who used AI effectively and those who became dependent on it. Our findings suggest this gap may be partly motivational. The students who described growing dependency on AI still expressed various forms of motivation~\cite{eccles2020expectancy}. Students across both patterns described reasons for valuing learning. They differed in whether and how they acted on those values when a faster alternative was available.

Achievement goal orientation theory offers one lens for understanding this difference. Elliot and McGregor's~\cite{elliot20012} framework distinguishes mastery-approach goals, focused on learning and developing competence, from performance-approach goals, focused on demonstrating competence relative to others. Critically, these orientations are not mutually exclusive. Students can hold both simultaneously. Our participants' descriptions are consistent with this pattern. Students like P3 and P12, who described effort as central to their identity and restricted their AI use accordingly, appeared to act primarily from a mastery orientation. Students like P10 and P13, who also expressed genuine interest in learning but whose behavior reflected a focus on grades and task completion, appeared to act from coexisting mastery and performance orientations, where performance goals pulled toward expedient AI use even as mastery goals valued effort. The learning-expedience tension our participants described may reflect this coexistence, a conflict within individual students between learning-oriented and completion-oriented goals. 

The seven students in our study who described putting in effort as a personal value (i.e., attainment value in SEVT terms) all managed this tension by maintaining alignment between their stated commitments and their behavior. Students who described behavior conflicting with their stated values did not always act on those values when a faster alternative was available, but they did mention valuing learning. This is consistent with Bastani et al.'s~\cite{bastani2024generative} finding that students who lost AI access after having it performed worse on assessments than those who never had it, suggesting that dependency patterns can form even among students who value learning. What appeared to distinguish students in our data was not stronger values but specific strategies, such as reframing effort around the learning process, drawing explicit boundaries around AI use, and tying those boundaries to personal identity.

\subsection{Implications for Teaching}

Our findings suggest three directions for instructional practice. Given that our study participants were engineering majors, we make note of where each implication would generalize broadly across introductory programming contexts, and where it is specific to a non-CS major population.

\subsubsection{Teach students to use AI instrumentally rather than expediently} Not all AI use in our study appeared harmful to motivation. Two students described AI as reducing frustration, and several who managed the tension used AI in bounded ways, such as checking syntax or verifying solutions after independent attempts. These patterns align with Nelson-Le Gall's~\cite{nelson1986help} distinction between instrumental help-seeking, which supports learning, and executive help-seeking, which bypasses it. Courses could explicitly teach this distinction and help students recognize when their AI use shifts from one to the other. Clear per-assignment AI policies appear to support this. Students who managed the tension described course policy as providing helpful structure for their decisions about when and how to use AI. Tiered policies that specify permitted levels of AI use for each assignment, which several students cited as helpful, offer one mechanism for encouraging instrumental use.

\subsubsection{Make the learning process visible and accountable} Knowledge that peers could use AI without consequences was a consistent source of frustration among students who maintained independent effort. Our social comparison findings suggest that enforcement mechanisms matter not only for compliance but for motivation. When students believe unauthorized AI use will be detected, the perceived cost of independent work may feel more justified. Process tracking, such as requiring intermediate submissions or monitoring problem-solving steps, can make the learning process itself valued rather than only the final artifact. Kazemitabaar et al.~\cite{kazemitabaar2025exploring} proposed ``Friction-Induced AI'' as a design approach that introduces deliberate friction to support cognitive engagement. This aligns with what some students in our study described doing voluntarily, such as restricting AI use to specific stages of their workflow. Course and tool designs that build this friction in may help students who lack the motivational anchors to create it on their own. 

One common institutional response to AI has been to shift assessment weight from practice assignments to proctored exams~\cite{lau2023ban}. While the shift to proctored exams addresses concerns about unauthorized AI use, it can demotivate students from completing homework assignments, which in turn can impact exam preparation. As P12 puts it: ``Because people aren't doing the homeworks as much, which means they're not learning as much. And because they're not learning as much, they're stressing even more about tests."
This matters especially for engineering students, for whom take-home programming assignments may be as important as proctored exams, since these assignments more closely resemble the work engineers will actually do. Additionally, these assignments give students room for the kind of productive struggle some described as central to their learning. As P3 frames it, ``the point is to struggle, the point is to learn.'' This pattern was broader than P3 alone, as other students described persisting through programming difficulty even when frustrated. Designs that monitor and value the process, not just the product, may preserve this struggle even when AI can produce the artifact.

\subsubsection{Foster existing sources of motivation} Students who described genuine enjoyment of programming appeared less likely to use AI expediently. Pedagogical approaches that cultivate intrinsic value, such as media computation~\cite{guzdial2003media} or personally relevant projects, may strengthen this buffer. For engineering students, contextualizing computing within authentic engineering problem-solving offers a parallel approach. Fostering a mastery goal orientation~\cite{elliot20012} and helping students see the career relevance of programming skills (utility value) may also help sustain effort. What counts as career-relevant, however, can feel different across contexts. Engineering students in our sample weighed programming against calculus and physics (P10, P11). All students face such trade-offs, but they may feel different when students take programming for another major rather than majoring in it. Utility-value work for engineering students may therefore need to anchor in how programming will serve their disciplinary engineering work, as P8 and P13 already did when framing programming as essential to complex calculations and MATLAB as ``an industry standard.'' Attainment value may serve as a strong anchor. The seven students in our sample who tied effort to personal identity managed the tension most consistently. For engineering students, the practical lever may be connecting programming effort to engineering identity, rather than attempting to recruit a new programmer identity. 

\section{Limitations}
\label{sec:limitations}

We note several limitations that bound what we can claim from this work.

First, our data are entirely self-reported. Students described their motivation and how AI availability shaped it, but we cannot verify whether their reported behavior matched their actual behavior or establish that AI availability caused the motivational patterns they described. 

Second, this study draws on a single course context, an introductory MATLAB course for engineering majors at one large public research university during one semester. Engineering students who program as a means to an end may relate to AI differently than CS majors for whom programming is a professional identity. The course's AI usage policy and tools (see Section~\ref{sec:methodology}) also shaped the context, and courses with different AI policies may produce different dynamics. Our findings should not be assumed to generalize without replication. 

Third, our design is cross-sectional. Each student was interviewed once during the final weeks of the semester. We captured a snapshot of how students described their motivation at that point, but we cannot determine whether these patterns represent stable orientations or a moment in an evolving process. Students' relationships with AI tools may shift as the tools become more capable or as students gain more programming experience. Longitudinal designs would be needed to distinguish trajectories from snapshots. 

Fourth, our interview protocol included a brief ($\sim$60 second) video demonstration of AI's programming capabilities to establish a shared reference point, followed by SEVT-structured questions that asked about each dimension first in general terms and then specifically in relation to AI. This ordering may have primed students to think about AI even when describing their baseline motivation, and the baseline-then-AI structure may have encouraged contrast effects that amplified perceived changes in either direction. In practice, students' AI-specific responses drew on their existing course experiences with AI tools rather than the video, and their responses ranged widely, from viewing AI as threatening to viewing it as beneficial, suggesting the protocol did not uniformly push responses in one direction. 

Fifth, our sample of 13 students is appropriate for qualitative thematic analysis but does not support claims about prevalence. Because recruitment materials mentioned AI, students who participated may have been more interested in programming or more reflective about their learning than the broader course population. To broaden participation, the instructor offered extra credit and we emphasized that participation was voluntary and confidential. That five students openly described behavior they viewed as conflicting with their stated values partially mitigates social desirability concerns. Still, students less reflective about AI's role may be underrepresented. 

Finally, students had access to both a course-specific AI assistant designed to scaffold learning and general-purpose tools such as ChatGPT. Our interview data do not always distinguish which tool students were referring to. Because these tools differ in design and capability, this ambiguity limits the specificity of our findings regarding particular tool designs. Our qualitative approach is appropriate for the questions we asked, as our goal was not to quantify motivational change but to understand how students reason about effort and motivation when AI is available. The patterns we identified provide a theoretical foundation that future work can test at scale, across contexts, and over time.

\section{Conclusion}

We investigated how 13 engineering students in an introductory MATLAB course described their motivation to put effort into programming assignments in the context of AI availability. Using Situated Expectancy-Value Theory, we found that AI availability entered into students' motivational reasoning primarily through cost, making the effort of independent work harder to justify rather than making the work itself harder. Nearly all students expressed a preference for learning through effort, but they diverged in whether they could act on those values. Students whose motivation was anchored in identity were better able to resist using AI expediently, while others described using AI in ways they viewed as undermining their own learning, even in a course with thoughtful AI policies. These findings suggest that the field's response to AI in computing education may benefit from attention to students' motivational structure, not only tool design and policy. We hope this work provides a foundation for motivation-aware approaches to supporting students as they learn to program alongside AI.

\begin{acks}
This material is based upon work supported by the National Science Foundation Graduate Research Fellowship Program under Grant No. DGE-2137100. Any opinions, findings, and conclusions or recommendations expressed in this material are those of the author(s) and do not necessarily reflect the views of the National Science Foundation.
\end{acks}

\bibliographystyle{ACM-Reference-Format}
\bibliography{main}

\end{document}